\documentclass{elsart}

\usepackage{framed,multirow}
\usepackage{url}
\usepackage{xcolor}

\usepackage{hyperref}

\usepackage{latexsym}
\usepackage{amssymb}
\usepackage{amsfonts}
\usepackage{amsmath,xcolor}
\usepackage{graphicx}
\usepackage{bm}

\def\vb#1{\vec{\mathbf{#1}}}
\def\ber{\begin{eqnarray}}
\def\eer{\end{eqnarray}}
\def\beq{\begin{equation}}
\def\eeq{\end{equation}}

\usepackage{natbib}

\usepackage{epsfig}

\usepackage{amssymb}

\begin{document}

\begin{frontmatter}



\title{Geometric definition of emission coordinates}


\author[a,b]{Matteo Luca Ruggiero \corauthref{cor}}
\corauth[cor]{Corresponding author}
\ead{matteo.ruggiero@polito.it}
\author[c,d]{Angelo Tartaglia}
\ead{angelo.tartaglia@polito.it}
\author[a]{Lorenzo Casalino}
\ead{lorenzo.casalino@polito.it}

\address[a]{DIMEAS, Politecnico di Torino, Corso Duca degli Abruzzi, 24, I-10129 Torino, Italy}
\address[b]{INFN - Legnaro, Via dell'Universit\`a 2, I-35020 Legnaro, Italy}
\address[c]{DISAT, Politecnico di Torino, Corso Duca degli Abruzzi, 24, I-10129 Torino, Italy}
\address[d]{INAF - OATo, Via Osservatorio 20, I-10025 Pino Torinese, Italy}


\begin{abstract}
We investigate a relativistic positioning system where the coordinates
of the users are determined by the proper times broadcasted by clocks in motion in spacetime: these are the so-called \textit{emission coordinates}. In particular, we focus on emission coordinates in flat spacetime: in this case, we show that these coordinates can be defined
using an approach  based on simple geometrical properties of geodesic triangles. We analyze the 2-dimensional case and then we show how the whole procedure can be applied to 4 dimensions, also in terms of ordinary three-dimensional spheres. An analytic solution for the coordinates of the receiver is obtained. The solution remains valid at almost any place, in particular when redundancy in the number of emitters can be exploited.
\end{abstract}

\begin{keyword}
relativistic positioning  \sep GNSS \sep GPS \sep Galileo
\end{keyword}

\end{frontmatter}


\section{Introduction} \label{sec:intro}

Current  positioning systems, such as GPS and Galileo  are essentially conceived as Newtonian, hence
based on a classical (i.e. Euclidean) space and absolute time, over which relativistic corrections are
added,  in order to take into account the effects arising from both the Special and General Theory of Relativity
(see \cite{ashby2003relativity,Ashby2004} and references therein). These systems involve the Earth and a constellation of satellites: they start from a terrestrial, non relativistic coordinate system and use the satellites  as moving beacons to indicate to the users their position. {Therefore, the signals emitted by the satellites cannot be used to define primary spacetime coordinates, since they have not enough information to relate  them to any other coordinate system}.  Furthermore, the clocks on the emitters are continuously corrected to be synchronized with clocks on ground: thus, the broadcasted signal does not contain the proper time, but a modified time.

There is another approach to the problem of global positioning, which brings in a shift from the Newtonian viewpoint to a true relativistic framework; the basic ideas of this approach were described in seminal papers written by different authors, such as {\cite{bahder2001navigation,coll2001elements,Rovelli:2001my,Blagojevic:2001gt,Coll:2006gv,bartolome2013relativistic}}, which shared a new and operational definition of spacetime coordinates. The starting assumption in the construction of such a new system of coordinates is that an ideal electromagnetic signal propagates along a null geodesic, and the main idea can be summarized as follows. Let us consider 4
clocks, moving along arbitrary world-lines in spacetime, and broadcasting their proper times, by means of
electromagnetic signals; then, any observer,  at a given spacetime point $P$  along his own world-line,
receives 4 numbers, carried by the 4 signals emitted by the clocks. These 4 numbers, say $(\tau_1,\tau_2,\tau_3,\tau_4)$, are nothing but the proper times of the emitting clocks and  constitute  {\textit{the coordinates of that spacetime point $P$ with respect to the system of emitters}}; they are usually referred to as \textit{emission coordinates}. In other words, the past light cone of a spacetime point $P$  cuts the world-lines of the clocks at 4 points:  the proper times measured  along these world-lines are the coordinates of
$P$. In practice, the clocks may either be supposed to be carried by satellites orbiting the Earth or at rest on the surface of our planet  (in any case, in what follows, they are referred to as \textit{emitters}), and the observers are the \textit{users} on the Earth, in the first case, or on board of
other satellites, in the second.

Such a system of coordinates can be considered as primary with respect to the spacetime structure if the
world-lines of the emitters are known; {the idea of a primary coordinate system was introduced by \citet{coll2001elements}}.
This is accomplished forcing for instance the emitters to
follow prescribed spacetime trajectories, which may be geodesics, if they are in free fall, or not.  
For all practical purposes and whenever the receiver wants to know its position with respect to other objects, the emission coordinates have to be converted into some other coordinate system encompassing the whole spacetime portion where measurements, exchange of signals and information and so on are taking place.

Of course we know that, under a few simple conditions of differentiability and absence of singularities, coordinates transformations are always possible in relativity: the real world does not depend on the coordinates you use, but your understanding of what is going on may be facilitated by an appropriate choice.  {The emission coordinates are therefore turned on a new and more practical reference frame}. What is however immediately clear in the case we are discussing is that in order to have a clear and univocal conversion you must know: a) where is the origin of the proper time for each emitter along its worldline; b) what are the space coordinates of the origins in the new reference frame; c) what are the equations of the worldlines of each emitter in the practical reference frame. The implementation of conditions a) and b) may be simpler if the emitters are at rest with respect to one another and their clocks synchronous in the new reference frame. Of course this cannot be the case when the emitters are different satellites orbiting a central mass.

A 2-dimensional approach to the new paradigm of relativistic positioning based on emission coordinates  was described by
\cite{Coll:2006nz,Coll:2006wk,Ruggiero:2007wh}, and its geometric nature was investigated by \cite{Coll:2006sg}: in this case two emitters are needed to define the emission coordinates of each event in spacetime. An explicit transformation between emission coordinates and inertial coordinates in flat spacetime was obtained by \cite{colproc,Coll:2009tm}.  Emission coordinates in a small region around the world-line of an observer in Schwarzschild spacetime were defined using Fermi coordinates by \cite{Bini:2008xw},
while the role of the gravitational perturbations in actual positioning systems around the Earth was considered by {\cite{delva2011numerical,gomboc2013relativistic,Puchades:2014oua,colmenero2021relativistic}}. A discrete relativistic positioning system was described by \cite{Carloni:2018cuf}. In addition, relativistic positioning systems were studied for space navigation, using signals from pulsars (see e.g. \cite{Ruggiero:2010nd,Tartaglia:2010sw,Tartaglia:2011ae,Bunandar:2011kp}).

In this paper, we focus on emission coordinates systems in flat spacetime, and show how they can be defined
using an approach  based on simple geometrical properties of geodesic triangles. We analyze the
2-dimensional case, that allows a simple and explicit analytic solution, and then we show how the whole procedure can be
applied also to the 4-dimensional spacetime.

\section{Geodesic Triangles in 2-dimensional spacetime and emission coordinates} \label{sec:geotria}



\begin{figure}[ht]
\begin{center}
\includegraphics[width=5cm,height=5cm]{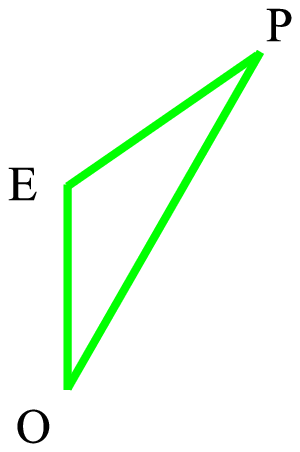}
\includegraphics[width=5cm,height=5cm]{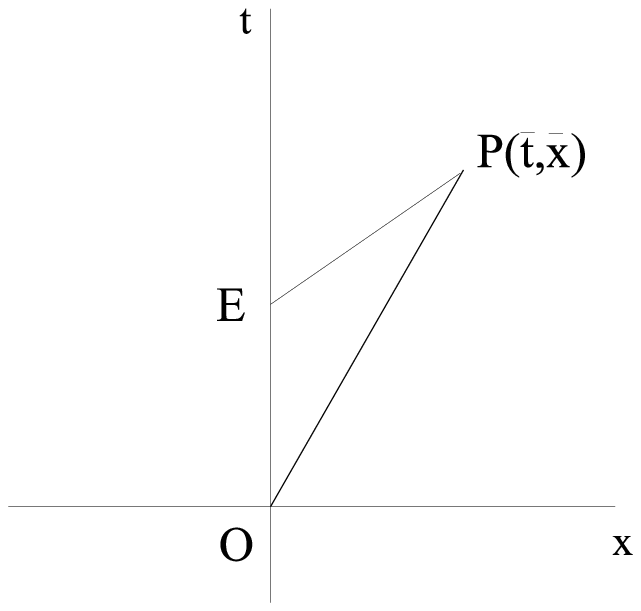}
\caption{Left: the geodesic triangle $OPE$. Right: an emitter is at rest in the reference system, where Cartesian
coordinates $t,x$ are used; its world-line is $\bm{x}(\tau)=\tau \bm{W}$, where $\bm{W} \equiv \left(1,0
\right)$ and $\tau$ is the proper time; the coordinates of point $P$ are such that $\overline{OP}=\bm{X}
\equiv \left(\bar t, \bar x\right)$, and its past light cone intersects the emitter's world-line at $E$.}
\label{fig:app12}
\end{center}
\end{figure}

It is well known that trigonometric relations can be obtained in Minkowski spacetime (see e.g. \cite{pascual,boccaletti2007space}). For instance, let us consider  a geodesic triangle in flat spacetime, i.e. a triangle whose sides are straight
lines (see figure \ref{fig:app12}, Left). Let the vertices be the \textit{events} or \textit{spacetime points}  $OPE$.\footnote{Greek indices
run from 0 to 3, lower case Latin indices run from 1 to 3; upper case Latin indices run from 1 to 4;  the
spacetime metric has signature $(-1,1,1,1)$, and we use units such that c=1;  boldface arrowed letters like
$\vb{k}$ refer to three-dimensional vectors; boldface letters like $\bm{x}$ refer to 4-dimensional vectors;
given two events in flat spacetime $O$ and $P$, then  $\overline{OP}$ is a {geodesic segment.}} Then, the following relation among the sides of the triangle holds (see
\cite{synge1960relativity}):

\beq |\overline{PE}|^2=|\overline{OP}|^2+|\overline{OE}|^2-2\overline{OP} \cdot \overline{OE},
\label{eq:triang1}
\eeq
which can be thought of as a generalization of the \textit{law of cosines}. We point out
that in Eq. (\ref{eq:triang1}) ``$|\ |^2$'' refers to the squared spacetime interval
between the two events,  and that ``$\cdot$'' means scalar product with respect to spacetime metric.
Moreover, we notice that Eq. (\ref{eq:triang1}) applies whatever the nature of the segment of the
triangle is (i.e. space-like, time-like, and null). {Eventually, we notice that Eq. (\ref{eq:triang1}) can be obtained also in curved spacetime for geodesic triangles, and correction terms appear (see, again,
\cite{synge1960relativity}).}

Now, let us show  how relation (\ref{eq:triang1}) can be used to introduce emission coordinates and, to this end,
we consider an emitter, moving with constant speed in  flat 2-dimensional spacetime. Without loss of
generality, we may choose a reference frame such that, in this frame, the emitter's world-line is $\bm{x}(\tau)=
\tau \bm{W}$, where  $\bm W$ is the unit vector $\bm{W} \equiv \left(1,0 \right)$ and $\tau$ is the emitter's proper time. In other words,
this emitter is at rest in the given reference frame. We consider an event $P$, whose coordinates are $\bm{X}
\equiv \left(\bar t, \bar x \right)$: we want to calculate the proper time $\tau$ at the intersection with the
past light cone from $P$. To this end, we apply the relation (\ref{eq:triang1}) to the triangle $OPE$,
where

\beq \overline{OP}=\bm{X}\equiv \left(\bar t, \bar x \right), \quad \overline{OE}=\tau \bm W \equiv \tau
\left(1,0 \right) \label{eq:defsides1}.
\eeq

 {Then, since the remaining side  is null, $|\overline{PE}|^{2}=0$ (see also figure \ref{fig:app12}, Right) and  Eq.
(\ref{eq:triang1}) reads:}

\beq 0=|\bm{X}|^2-\tau^2-2\tau \bm{X}\cdot\bm{W}. \label{eq:triang2} \eeq

Hence, we obtain\footnote{The sign has been chosen in order to select the intersection with the past light cone of $P$.  {Remember also that with the chosen signature the scalar product between two timelike vectors is negative.} {Eventually, notice that when solving the square root in (\ref{eq:triang3}) it is relevant to consider the direction of propagation of the signals.}},

\beq \tau= -\bm{X}\cdot\bm{W} - \sqrt{\left(\bm{X}\cdot\bm{W} \right)^2+|\bm{X}|^2}. \label{eq:triang3}
\eeq

Eq. (\ref{eq:triang3}) was obtained in previous papers (e.g. \cite{Rovelli:2001my} and \cite{Blagojevic:2001gt}) and it is manifestly
Lorentz-covariant: as a consequence, it is true in any Lorentz reference frame.

\begin{figure}[t]
\begin{center}
\includegraphics[width=5cm,height=5cm]{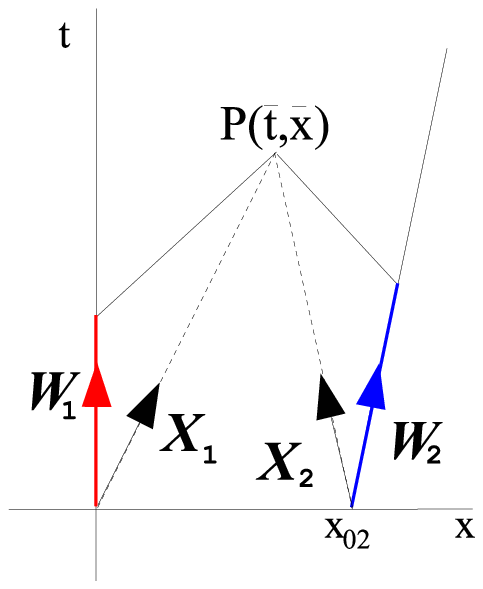}
\includegraphics[width=5cm,height=5cm]{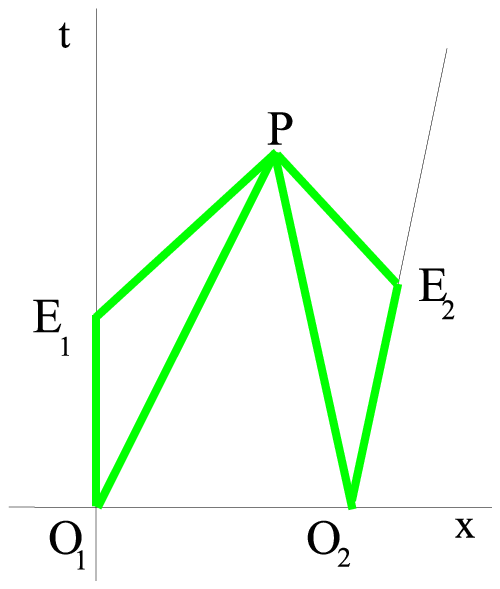}
\caption{Left: An emitter is at rest, so that its world-line is parallel to the unit vector $\bm W_1$, while
another emitter is moving and its world-line is parallel to the unit vector $\bm W_2$. The Cartesian coordinates
of the event $P$ can be expressed in terms of the proper times along the world-lines of the emitters. Right:
Considering the world-lines of the two emitters, and the point $P$, we can build the two  geodesic triangles
$O_1 E_1 P$ and $O_2 E_2 P$, where $E_1,E_2$ are the intersections of the past light cone of $P$ with the two
world-lines.} \label{fig:geo1}
\end{center}
\end{figure}

This relation can be used to define a map between  emission and Cartesian coordinates. To begin with, we refer again to a 2-dimensional spacetime model: in this case, \text{two} emitters are necessary, so we may write their world-lines (see figure
\ref{fig:geo1}, Left) in Minkowski plane in the general form:
\beq \label{eq:defgeo21} x^\alpha_A\left(\tau_A \right)=W^\alpha_A\tau_A+x_{0A}^\alpha\ , \qquad A=1,2.
\eeq
which can be written in vector form
\beq
\bm x_{A}(\tau_{A})=\tau_{A}\bm W_{A}+\bm x_{0A}. \label{eq:defgeo22}
\eeq

{For the sake of simplicity, we initially consider one of the emitters at rest in the given reference frame, and assume that the world-lines are such that}
\begin{eqnarray}
\bm{W}_1 &\equiv& \left(1,0 \right), \label{eq:defWl1} \\
\bm{W}_2 &\equiv& \gamma_2 \left(1, v_2 \right). \label{eq:defWm1}
\end{eqnarray}
\beq t_{01}=0\quad x_{01}=0,  \label{eq:defl1} \eeq \beq t_{02}=0\quad x_{02} \neq 0.  \label{eq:defl2} \eeq

In other words, the first emitter is at rest in the given reference frame, while the other one is moving with
constant linear velocity $v_2$ ($\displaystyle \gamma_2=\frac{1}{\sqrt{1-v_{2}^{2}}}$ is the corresponding Lorentz factor); $\tau_1,\tau_2$ are the proper
times along the two world-lines;  the parametrization is such that the  position of the second emitter at
$\tau_2=0$ is $x_{02}$. We consider an event  $P$ whose coordinates are $\left(\bar t,\bar x \right)$.
According to the approach described above, it is possible to define the proper times $\tau_1$ and $\tau_2$, as measured along the two world-lines, at the
intersections with the past light cone from $P$. This can be done by exploiting eq. (\ref{eq:triang3}),
and considering the geodesic triangles $O_1 E_1 P$ and $O_2 E_2 P$ (see figure \ref{fig:geo1}, Right), where

\beq \overline{O_1P }=\bm{X}_1 \equiv \left(\bar t, \bar x \right),  \quad \overline{O_1E_1}=\tau_1 \bm W_1
\equiv \tau_1 \left(1,0 \right), \label{eq:deftria1} \eeq \beq \overline{O_2P }=\bm{X}_2 \equiv  \left(\bar t,
\bar x-x_{02} \right), \quad \overline{O_2E_2}=\tau_2 \bm W_2 \equiv \tau_2 \gamma_2 \left(1, v_2 \right),
\label{eq:deftria2}
\eeq

and the remaining  sides $\overline{P E_1}$ and $\overline{P E_2}$ are null. On  applying eq. (\ref{eq:triang3})
to these triangles, we obtain the system of equations

\beq
\left\{\begin{array}{ccc}\tau_1 &=& -\bm{X}_1\cdot\bm{W}_1 - \sqrt{\left(\bm{X}_1\cdot\bm{W}_1 \right)^2+|\bm{X}_1| ^2} \\\tau_2&=& -\bm{X}_2\cdot\bm{W}_2 - \sqrt{\left(\bm{X}_2\cdot\bm{W}_2 \right)^2+|\bm{X}_2|^2} \end{array}\right. \label{eq:trianglm}
\eeq


which may be solved for $(\bar t, \bar x)$ as functions of $(\tau_1,\tau_2)$ thanks to Eqs.
(\ref{eq:defWl1})-(\ref{eq:defWm1}) and (\ref{eq:deftria1})-(\ref{eq:deftria2}):

\begin{eqnarray}
\bar t &=&\frac{1}{2}\left(\sqrt{\frac{1+v_2}{1-v_2}}\tau_2+x_{02}+\tau_1 \right), \label{eq:solTstau1} \\
\bar x &=&\frac{1}{2}\left(\sqrt{\frac{1+v_2}{1-v_2}}\tau_2+x_{02}-\tau_1 \right). \label{eq:solXstau1}
\end{eqnarray}

These relations define the Cartesian coordinates $\left(\bar t,\bar x \right)$ of the point P in terms of the emission coordinates $(\tau_1,\tau_2)$ and of the \textit{known} emitter parameters.\\
{The above procedure enables to define a map between Cartesian and emission coordinates.
It is important to notice that emission coordinates can be always obtained by the intersection of the past light-cone with the emitter world-lines; however, there are configurations     where the  map between Cartesian and emission coordinates is not bijective, for instance when one emitter is on the light-cone of the other(see e.g. \cite{Coll:2006gv,Coll:2006nz}): in fact, in the 2-dimensional case, a solution is possible only when the emitters do not lie on the same light cone, and in this case the receiver (point P) is between the emitters. The system (\ref{eq:trianglm}) is undetermined when the emitters are on the same light cone, as the signals travel in the same direction.}

Actually, emission coordinates can be used to write  the spacetime metric interval: {as showed by \cite{Coll:2006nz}}, thanks to the map defined by Eqs. (\ref{eq:solTstau1})-(\ref{eq:solXstau1}), the metric of Minkowski spacetime in terms of emission coordinates is
given by:
\begin{equation}
ds^2=-\sqrt{\frac{1+v_2}{1-v_2}}d\tau_1 d\tau_2, \label{eq:metricaflattaus1}
\end{equation}
which, introducing the shift $\sigma_{2}$
\beq \sigma_{2} \doteq \sqrt{\frac{1+v_{2}}{1-v_{2}}} \label{eq:deflambda1}\eeq
for an emitter moving with constant linear velocity $v_{2}$, can be written as
\begin{equation}
ds^2=-\sigma_2 d\tau_1 d\tau_2. \label{eq:metricaflattaus1lambda2}
\end{equation}

\begin{figure}[t]
\begin{center}
\includegraphics[width=4cm,height=4cm]{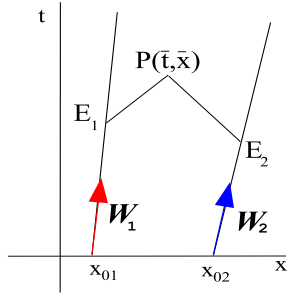}
\caption{{The two emitters are in motion with linear velocities $v_1,v_2$, both positive in this sketch. In the situation considered, the spatial location  of the first emitter precedes that of the second one.} } \label{fig:geo1ter}
\end{center}
\end{figure}

It is easy to check that if we had considered also the first emitter in motion with linear velocity $v_1$ and
initial position $x_{01}$ {(see Figure \ref{fig:geo1ter})} we would have obtained the following results:

\begin{eqnarray}
\bar t &=&\frac{1}{2}\left(\sqrt{\frac{1-v_1}{1+v_1}}\tau_1+\sqrt{\frac{1+v_2}{1-v_2}}\tau_2+x_{02}-x_{01} \right), \label{eq:solTstau2} \\
\bar x &=&\frac{1}{2}\left(-\sqrt{\frac{1-v_1}{1+v_1}}\tau_1+\sqrt{\frac{1+v_2}{1-v_2}}\tau_2+x_{02}+x_{01}
\right). \label{eq:solXstau2}
\end{eqnarray}
{Notice that the above expressions are not symmetric with respect to the two emitters: the signal from $E_1$ travels towards positive x and the signal from $E_2$ in the opposite direction. }

The metric of Minkowski spacetime in this case, turns out to be

\begin{equation}
ds^2=-\frac{\sigma_2}{\sigma_1} d\tau_1 d\tau_2. \label{eq:metricaflattaus1lambda12}
\end{equation}
In other words, the metric function is constant and it equals the relative shift between the two emitters.
Indeed, these results are in agreement with those obtained in  \cite{Coll:2006nz}.

{We end this Section by reformulating the system of equations (\ref{eq:trianglm}) using a different notation.  The emitter world-lines are expressed in the general form (\ref{eq:defgeo22}), where we notice that $\bm x_{0A}$ is the event along the world-line corresponding to zero proper time. So, if we consider two arbitrary geodesic emitters and denote by $\bm{X}\equiv \left(\bar t, \bar x \right)$ the position of the receiver, we obtain the following equations
\beq
\left\{\begin{array}{ccc}\bm X \cdot \bm X -2 \bm X \cdot \left(\bm x_{01}+\tau_{1} \bm W_{1} \right)-\tau_{1}^{2}+\bm x_{01}\cdot \bm x_{01}+2  \tau_{1}\bm x_{01}\cdot \bm W_{1 }&=&0,  \\ \bm X \cdot \bm X -2 \bm X \cdot \left(\bm x_{02}+\tau_{2} \bm W_{2} \right)-\tau_{2}^{2}+\bm x_{02}\cdot \bm x_{02}+2  \tau_{2}\bm x_{02}\cdot \bm W_{2 }&=&0 \end{array}\right. \label{eq:sys2D00}
\eeq
According to the geodesic equation (\ref{eq:defgeo22}), it is $\bm x_{i}=\tau_{i} \bm W_{i}+\bm x_{0i}$, and we define  $p_{i}=-\tau_{i}^{2}+\bm x_{0i}\cdot \bm x_{0i}+2  \tau_{i}\bm x_{0i}\cdot \bm W_{i}$, where $i=1,2$. Then, the above system becomes
\beq
\left\{\begin{array}{ccc}\bm X \cdot \bm X-2\bm X \cdot \bm x_{1}+p_{1}&=&0 \\ \bm X \cdot \bm X-2\bm X \cdot \bm x_{2}+p_{2}&=&0 \end{array}\right. \label{eq:sys2D000}
\eeq
We notice that both the vectors $\bm x_{i}$ and the scalars $p_{i}$ contain only \textit{known information}, that is the parameters of the world-lines and the proper times measured along them. As we are going to show in next Section for the 4-dimensional case, from the system in the form (\ref{eq:sys2D000})  it is possible to obtain an analytical solution.}

\section{The 4-dimensional case} \label{sec:4D}


Let us consider  the general  4-dimensional case: first, we want to show how emission coordinates can be obtained from  emitters
moving with constant linear velocities in the 4-dimensional Minkowski spacetime,  exploiting the properties of
geodesic triangles. Then, we will show how we can define a map between emission coordinates and Cartesian coordinates.

Indeed, the situation depicted in figure (\ref{fig:geo1}) can be  generalized to the
4-dimensional spacetime. Let us consider the spacetime point $P$, having coordinates $(\bar t,\bar x, \bar y,
\bar z)$, so that $P=P(\bar t,\bar x, \bar y, \bar z)$, and the corresponding {4-vector} is
\beq
\overline{OP}= \bm X \equiv \left(\bar t,\bar x, \bar y, \bar z\right). \label{eq:defX4d1} \eeq
The emitters
world-lines are geodesics of flat spacetime, i.e. straight lines, that can be written in the form (\ref{eq:defgeo22})
with normalized 4-velocities \beq \bm W_A=\gamma_A \left(1,v_A \vb n_A \right), \quad A=1,2,3,4, \label{eq:defWA4d1}
\eeq
parameterized by the linear velocity $v_A$, with the corresponding Lorentz factor $\displaystyle \gamma_A=\frac{1}{\sqrt{1-v_{A}^{2}}}$, and where $\vb
n_A$ denotes a three-dimensional unit vector representing the spatial direction of motion. Let us also define
the vectors
\beq \overline{OO_A}=\bm x_{0A}\equiv\left(t_{0A},x_{0A},y_{0A},z_{0A} \right), \quad A=1,2,3,4,
\label{eq:defX0A4d1} \eeq which determine the positions at $\tau_A=0$ of the 4 emitters, and
\beq
\overline{O_{A}P}=\bm X-\bm x_{OA}\equiv\left(\bar t-t_{0A},\bar x-x_{0A},\bar y-y_{0A},\bar z-z_{0A} \right)
\label{OAP}
\eeq
which define the position of the event $\bm X$ with respect to the emitters position at $\tau_{A}=0$. We suppose to know the emitters world-lines, which means that both vectors $\bm W_{A}$ and $\bm x_{0A}$ are known.

\begin{figure}[ht]
\begin{center}
\includegraphics[width=6cm,height=6cm]{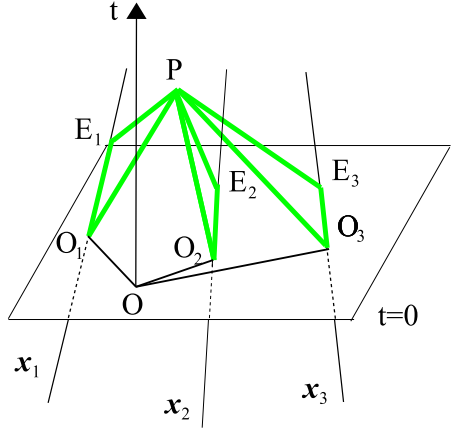}
\caption{{For the  A-th emitter, $O_A$ is its position (in spacetime) at $t=0$ and the event $E_A$ corresponds to
the intersection between the past light cone at $P$ and the emitter's world-line. Then, the geodesic
triangle $O_AE_AP$ is naturally defined.}} \label{fig:geo4}
\end{center}
\end{figure}

To visualise the situation we consider  a 3-dimensional spacetime, such as in Figure \ref{fig:geo4} and, for the sake of simplicity,  we suppose that $t_{0A}=0$: in other words, the emitters proper times are zero when $t=0$. Then, the
vectors $\bm X_{0A}$ describe the emitters' position at $t=0$ (see figure \ref{fig:geo4}) and the geodesics triangles
$O_AE_AP$ are naturally defined.  The application to these triangles of our basic relation (\ref{eq:triang1}), gives
\beq |\overline{PE_{A}}|^2=|\overline{O_{A}P}|^2+|\overline{O_{A}E_{A}}|^2-2\overline{O_{A}P} \cdot \overline{O_{A}E_{A}},
\label{eq:triang33}
\eeq

Recalling that $\overline{PE_{A}}$ is null, it is easy to check that Eq. (\ref{eq:triang33}) is equivalent to
\beq
\left|\bm X-\bm x_{A} \right|^{2}=0  \rightarrow  \left|(\bm X-\bm x_{0A})-\tau_{A}\bm W_{A} \right|^{2}=0  \quad A=1,2,3,4, \label{eq:lightcone}
\eeq
in other words,  the past light cone of P intersects the emitters world-lines at the events $E_{A}$.

Consequently, using the above definitions (\ref{eq:defgeo22}) and (\ref{eq:defX0A4d1}), from (\ref{eq:lightcone}) we explicitly obtain
\beq
\bm X \cdot \bm X -2 \bm X \cdot \left(\bm x_{0A}+\tau_{A} \bm W_{A} \right)-\tau_{A}^{2}+\bm x_{0A}\cdot \bm x_{0A}+2  \tau_{A}\bm x_{0A}\cdot \bm W_{A }=0 \label{eq:ex1A}
\eeq

{If we use the geodesic equation (\ref{eq:defgeo22}), we may set  $\displaystyle \bm x_{A}=\tau_{A} \bm W_{A}+\bm x_{0A}$ and $p_{A}=-\tau_{A}^{2}+\bm x_{0A}\cdot \bm x_{0A}+2  \tau_{A}\bm x_{0A}\cdot \bm W_{A }$, then the above equation becomes
\beq
\bm X \cdot \bm X-2\bm X \cdot \bm x_{A}+p_{A}=0. \label{eq:ex2A}
\eeq
As we noticed in the previous Section,  $\bm x_{A}$ and $p_{A}$ contain only \textit{known} information. Hence, if we take into account the 4 emitters, we may write the system
\beq
\left\{\begin{array}{ccc} 0 & = & \bm X \cdot \bm X-2\bm X \cdot \bm x_{1}+p_{1}  \\ 0 & = & \bm X \cdot \bm X-2\bm X \cdot \bm x_{2}+p_{2}  \\ 0 & = & \bm X \cdot \bm X-2\bm X \cdot \bm x_{3}+p_{3} \\ 0 & = & \bm X \cdot \bm X-2\bm X \cdot \bm x_{4}+p_{4} \end{array}\right. \label{eq:exsys1}
\eeq
which can be solved for $\bm X \equiv \left(\bar t,\bar x, \bar y, \bar z\right)$ knowing the emitters set of parameters $\bm x_{A}$ and $p_{A}$. In what follows, we describe two approaches to solving the system (\ref{eq:exsys1}).\\ \\
It is possible to obtain an analytical solution using the approach introduced by \cite{bancroft1985algebraic}. To this end, we are going to formally define suitable matrices and column vectors to rewrite the above system (\ref{eq:exsys1}) in a way that lends itself to a direct algebraic solution.\\
We define the  matrices
\beq
\mathcal{L}=\left(\begin{array}{cccc}x_{1}^{0} & x_{1}^{1} & x_{1}^{2} & x_{1}^{3} \\x_{2}^{0} & x_{2}^{1} & x_{2}^{2} & x_{2}^{3} \\x_{3}^{0} & x_{3}^{1} & x_{3}^{2} & x_{3}^{3} \\x_{4}^{0} & x_{4}^{1} & x_{4}^{2} & x_{4}^{3}\end{array}\right), \quad \mathcal{M}=\left(\begin{array}{cccc}-1 & 0 & 0 & 0 \\0 & 1 & 0 & 0 \\0 & 0 & 1 & 0 \\0 & 0 & 0 & 1\end{array}\right)  \label{eq:defLM}
\eeq
where, in $x_{A}^{\alpha}$, $A$ refers to the emitter and $\alpha$ to its spacetime components, and the column vectors
\beq
\bm{\mathcal{A}}=\left(\begin{array}{c}\frac 1 2 p_{1} \\ \frac 1 2 p_{2} \\ \frac 1 2 p_{3} \\ \frac 1 2 p_{4}\end{array}\right), \quad \bm{\mathcal{I}}=\left(\begin{array}{c}1 \\1 \\1 \\1\end{array}\right). \label{eq:defAIX}
\eeq}
Eventually, let's define
\beq
\lambda=\frac 1 2 \bm{{X}} \cdot \bm{{X}},\label{eq:deflambda11}
\eeq
{We remember that the Minkowski functional (scalar product) between any pairs of vectors in 4-space
\beq
 \bm{a}=\left(\begin{array}{c}a^{0} \\ a^{1} \\ a^{2} \\ a^{3}\end{array}\right), \quad \bm{b}=\left(\begin{array}{c}b^{0} \\ b^{1} \\ b^{2} \\ b^{3}\end{array}\right) \label{eq:4vectors}
\eeq
is defined by
\beq
\bm{a} \cdot \bm{b}=a^{1}b^{1}+a^{2}b^{2}+a^{3}b^{3}-a^{0}b^{0}  \label{eq:4vectors1}
\eeq}
Then, the above system (\ref{eq:exsys1}) can be written in the form
\beq
\lambda \bm{\mathcal{I}}-\mathcal L \mathcal M \bm{ X}+\bm{\mathcal A} =0, \label{eq:sys1}
\eeq
whose solution is formally written as
\beq
\bm{ X} =\mathcal{M} \mathcal{L}^{-1}\left(\lambda \bm{\mathcal{I}}+\bm{\mathcal{A}} \right) \label{eq:sys2}
\eeq
Then, using the properties of the scalar product we obtain the following equation
\beq
e\lambda^{2}+2f\lambda+g=0  \label{eq:sys3}
\eeq
where
\beq
e=\left(\mathcal{L}^{-1}\bm{\mathcal{I}}\right) \cdot  \left( \mathcal{L}^{-1}\bm{\mathcal{I}}\right), \quad f=\left(\mathcal{L}^{-1}\bm{\mathcal{I}}\right) \cdot  \left( \mathcal{L}^{-1}\bm{\mathcal{A}}\right)-1, \quad g=\left(\mathcal{L}^{-1}\bm{\mathcal{A}}\right) \cdot  \left( \mathcal{L}^{-1}\bm{\mathcal{A}}\right) \label{eq:sys4}
\eeq
Eq. (\ref{eq:sys3}) can be solved for $\lambda$ in terms of known parameters, then substituting in (\ref{eq:sys2}) we obtain the solution for the coordinates. Notice that there are in general two solutions, which correspond to past and future light cone intersections.

{We remark that, in order to solve the system (\ref{eq:exsys1}) with the above described procedure, the matrix $\mathcal L$ needs to be non singular: this amounts to saying that none of the vectors $\bm x_{A}$  is a linear combination of the others.\footnote{{These vectors define the positions of the emitters at the emission of the signals, with respect to the origin of the reference frame.  If a couple of these vectors are parallel to each other
this means that they lie on the same null world-line crossing $\bm X$: as a consequence  $\bm X$ is a null vector and $\lambda=0$ in Eq. (\ref{eq:sys1}). Then,  we have just three independent equations in the form (\ref{eq:exsys1}), that can be solved with the constraint $\bm X \cdot \bm X=0$. If more than two emission events are aligned, then the system cannot be solved.}}
As discussed by \citet{Coll:2009tm}, the map from emission coordinates to usual (e.g. Cartesian) coordinates is well defined only for suitable configuration of the emitters, which, for $t=0$, need to lie on a space-like hypersurface. As we notice before, this issue is present also in the 2-dimensional case. More generally, this is related with the bifurcation problem for positioning systems (see \cite{abel1991existence,chaffee1994exact,coll2012positioning}).\\
\indent One possible solution to avoid singular configurations is to exploit  redundancy, that is the possibility to consider more than 4 emitters. Redundancy is the basis of another approach to the positioning procedure which again leads to the formal solution for the coordinates of the receiver.}

Reference can be made to the future light cones of the emitters. Such light cones, when projected onto the space of the emitter at the time of emission, correspond to spheres centered on the emitter:
\beq{(x-x_A)^2+(y-y_A)^2+(z-z_A)^2=r_A^2}
\eeq
Now $x_A$ is a shorthand for the corresponding cartesian component of $\bm x_{A}$, see Eq. (\ref{eq:defgeo22}).

Considering the flatness of spacetime and the invariance of the speed of light, the typical radius of a sphere is $r_A=t-t_A$ where of course it is also $t_A=\gamma_A \tau_A$.

The equation of a typical sphere is now:
\beq{(t-t_A)^2=(x-x_A)^2+(y-y_A)^2+(z-z_A)^2} \label{sph}
\eeq

In order to get immediately rid of quadratic terms in the unknowns it is convenient to use one emitter more than the dimension of space time, i.e. in our case $5$ emitters (so that now it is $A=1,2,3,4,5$). We may then write a system of five equations like (\ref{sph}). Subtracting member to member pairs of such equations we come back to a system of four equations linear in the unknowns, i.e. in the coordinates of the receiver, which is located at the intersection of the five spheres. The typical equation of the new system is:
\begin{multline}
2(t_A-t_{A+1})t-2(x_A-x_{A+1})x-2(y_A-y_{A+1})y-2(z_A-z_{A+1})z= \\
x_{A+1}^2-x_A^2+y_{A+1}^2-y_A^2+z_{A+1}^2-z_A^2-(t_{A+1}^2-t_A^2) \label{eq:diff11}
\end{multline}
In compact matrix notation the system to be solved is now
\beq
\mathcal{C} \bm{ X}=\bm{\mathcal N}
\eeq


The known matrix and vector are:
\beq{\mathcal{C}=2\left(\begin{array}{cccc} t_1-t_2 & x_2 -x_1 & y_2-y_1 & z_2-z_1 \\ t_2-t_3 & x_3 -x_2 & y_3-y_2 & z_3-z_2 \\t_3-t_4 & x_4 -x_3 & y_4-y_3 & z_4-z_3 \\t_4-t_5 & x_5 -x_4 & y_5-y_4 & z_5-z_4\end{array}\right)  \label{eq:coef}}
\eeq
and
\beq{\bm{\mathcal N}=\left(\begin{array}{c} x_2^2-x_1^2+y_2^2-y_1^2+z_2^2-z_1^2+t_1^2-t_2^2 \\ x_3^2-x_2^2+y_3^2-y_2^2+z_3^2-z_2^2+t_2^2-t_3^2 \\x_4^2-x_3^2+y_4^2-y_3^2+z_4^2-z_3^2+t_3^2-t_4^2 \\x_5^2-x_4^2+y_5^2-y_4^2+z_5^2-z_4^2+t_4^2-t_5^2\end{array}\right)  \label{eq:know}}
\eeq

The solution is

\beq
 \bm{ X}=\mathcal{C}^{-1} \bm{\mathcal N}
\eeq


It is well defined and unique provided that the matrix $\mathcal{C}$ is non singular. {We remark that when the solution of the above system exists, it does not depend on the choice of the equations used to make the differences in Eq. (\ref{eq:diff11}) and build the matrix $\mathcal{C}$.}



In summary, the approach we have exposed enables to define a map between emission coordinates, and Cartesian coordinates, once the world-lines of the emitters are known and excluding a limited number of special cases we have mentioned.

\section{Conclusions} \label{sec:conc}

In the present paper we outlined the baseline computations to define a relativistic positioning system based on signals received by emitters.
Emission coordinates are defined by intersecting the past light cone of a given event with the world-lines of the emitters, starting from the general geometrical relation (\ref{eq:triang1}), applied to the geodesic
 motion of emitters. Emission coordinates are the proper times measured by clocks transported by emitters, and they are mapped into the usual inertial or Cartesian coordinates in an $n$-dimensional spacetime when the world lines of $n$ emitters are known. {The basic 2-dimensional case is extended to the more realistic 4-dimensional one, for which we obtained  analytical solutions for the coordinates starting from the signals of the emitters.}

We point out that a different procedure to obtain this map was introduced in comparison to \cite{Coll:2009tm,colproc}:
in particular, the authors consider arbitrary world lines and the system (\ref{eq:exsys1}) is split into a single quadratic equation and a degenerate linear system, since the unknowns are referred to a specific emitter.

{In flat spacetime, our approach can be generalised to arbitrary world-lines without difficulties: indeed starting from the light-cone equation (\ref{eq:lightcone}), the procedure can be extended once the expressions of the world-lines are known as function of the coordinates of the primary reference frame (which amounts to have $\tau$ as a function of those coordinates).
A special case of interest could be a situation where the emitters are at rest in a uniformly rotating frame, which could mimic the more realistic condition of emitters at rest on the surface of the Earth that can be used to positioning satellites around it. In the latter case, it is relevant to explore the possibility to extend our approach to curved spacetime: this will be done in future works. To this end, it is important to point out that, according to what is already known in the literature (see e.g. \cite{Puchades:2014oua}), for a practical situation of emitters moving around the Earth used to build a terrestrial positioning system,   the effect of the gravitational field of the Earth can be neglected, since the distance traveled by photons is small: in fact, errors due to the uncertainties in the knowledge of the emitters world lines are greater than those effects in any realistic situation which involves Galileo or GPS satellites.  }

\textbf{Acknowledgements.}\ This work is part of the G4S 2.0 project, developed under the auspices of the Italian Space Agency (ASI) within the frame of the Bando Premiale CI-COT-2018-085 with co-participation of the Italian Institute for Astrophysics (INAF) and the Politecnico di Torino (POLITO). The scientific research carried out for the project is supported under the Accordo Attuativo n. 2021-14-HH.0.


\bibliographystyle{elsarticle-harv}
\bibliography{LC}

\begin{thebibliography}{31}
\expandafter\ifx\csname natexlab\endcsname\relax\def\natexlab#1{#1}\fi
\expandafter\ifx\csname url\endcsname\relax
  \def\url#1{\texttt{#1}}\fi
\expandafter\ifx\csname urlprefix\endcsname\relax\def\urlprefix{URL }\fi

\bibitem[{Abel and Chaffee(1991)}]{abel1991existence}
Abel, J.~S., Chaffee, J.~W., 1991. Existence and uniqueness of gps solutions.
  IEEE Transactions on Aerospace and Electronic Systems 27~(6), 952--956.

\bibitem[{Ashby(2003)}]{ashby2003relativity}
Ashby, N., 2003. Relativity in the global positioning system. Living Reviews in
  relativity 6~(1), 1--42.

\bibitem[{Ashby(2004)}]{Ashby2004}
Ashby, N., 2004. The sagnac effect in the global positioning system. In: Rizzi,
  G., Ruggiero, M.~L. (Eds.), Relativity in Rotating Frames: Relativistic
  Physics in Rotating Reference Frames. Springer Netherlands, Dordrecht, pp.
  11--28.

\bibitem[{Bahder(2001)}]{bahder2001navigation}
Bahder, T.~B., 2001. Navigation in curved space--time. American Journal of
  Physics 69~(3), 315--321.

\bibitem[{Bancroft(1985)}]{bancroft1985algebraic}
Bancroft, S., 1985. An algebraic solution of the gps equations. IEEE
  transactions on aerospace and electronic systems AES-21~(1), 56--59.

\bibitem[{Bini et~al.(2008)Bini, Geralico, Ruggiero, and
  Tartaglia}]{Bini:2008xw}
Bini, D., Geralico, A., Ruggiero, M.~L., Tartaglia, A., 2008. {Emission versus
  Fermi coordinates: Applications to relativistic positioning systems}. Class.
  Quant. Grav. 25, 205011.

\bibitem[{Blagojevic et~al.(2002)Blagojevic, Garecki, Hehl, and
  Obukhov}]{Blagojevic:2001gt}
Blagojevic, M., Garecki, J., Hehl, F.~W., Obukhov, Y.~N., 2002. {Real null
  coframes in general relativity and GPS type coordinates}. Phys. Rev. D 65,
  044018.

\bibitem[{Boccaletti et~al.(2007)Boccaletti, Catoni, and
  Catoni}]{boccaletti2007space}
Boccaletti, D., Catoni, F., Catoni, V., 2007. Space-time trigonometry and
  formalization of the ``twin paradox'' for uniform and accelerated motions.
  Advances in applied Clifford algebras 17~(1), 1--22.

\bibitem[{Bunandar et~al.(2011)Bunandar, Caveny, and Matzner}]{Bunandar:2011kp}
Bunandar, D., Caveny, S.~A., Matzner, R.~A., 2011. {Measuring emission
  coordinates in a pulsar-based relativistic positioning system}. Phys. Rev. D
  84, 104005.

\bibitem[{Carloni et~al.(2020)Carloni, Fatibene, Ferraris, McLenaghan, and
  Pinto}]{Carloni:2018cuf}
Carloni, S., Fatibene, L., Ferraris, M., McLenaghan, R.~G., Pinto, P., 2020.
  {Discrete Relativistic Positioning Systems}. Gen. Rel. Grav. 52~(2), 12.

\bibitem[{Chaffee and Abel(1994)}]{chaffee1994exact}
Chaffee, J., Abel, J., 1994. On the exact solutions of pseudorange equations.
  IEEE Transactions on Aerospace and Electronic Systems 30~(4), 1021--1030.

\bibitem[{Coll(2001)}]{coll2001elements}
Coll, B., 2001. Elements for a theory of relativistic coordinate systems:
  formal and physical aspects. In: Reference Frames and Gravitomagnetism. World
  Scientific, pp. 53--65.

\bibitem[{Coll(2006)}]{Coll:2006gv}
Coll, B., 2006. {Relativistic positioning systems}. AIP Conf. Proc. 841~(1),
  277--284.

\bibitem[{Coll(2013)}]{bartolome2013relativistic}
Coll, B., 2013. Relativistic positioning systems: perspectives and prospects.
  Acta Futura 7, 35--47.

\bibitem[{Coll et~al.(2006{\natexlab{a}})Coll, Ferrando, and
  Morales}]{Coll:2006nz}
Coll, B., Ferrando, J.~J., Morales, J.~A., 2006{\natexlab{a}}. {A
  Two-dimensional approach to relativistic positioning systems}. Phys. Rev. D
  73, 084017.

\bibitem[{Coll et~al.(2006{\natexlab{b}})Coll, Ferrando, and
  Morales}]{Coll:2006wk}
Coll, B., Ferrando, J.~J., Morales, J.~A., 2006{\natexlab{b}}. {Positioning
  with stationary emitters in a two-dimensional space-time}. Phys. Rev. D 74,
  104003.

\bibitem[{{Coll} et~al.(2009){Coll}, {Ferrando}, and {Morales}}]{colproc}
{Coll}, B., {Ferrando}, J.~J., {Morales}, J.~A., May 2009. {Emission
  coordinates in Minkowski space-time}. In: {Kunze}, K.~E., {Mars}, M.,
  {V{\'a}zquez-Mozo}, M.~A. (Eds.), Physics and Mathematics of Gravitation.
  Vol. 1122 of American Institute of Physics Conference Series. pp. 225--228.

\bibitem[{Coll et~al.(2010)Coll, Ferrando, and Morales-Lladosa}]{Coll:2009tm}
Coll, B., Ferrando, J.~J., Morales-Lladosa, J.~A., 2010. {Positioning Systems
  in Minkowski Space-Time: from Emission to Inertial Coordinates}. Class.
  Quant. Grav. 27, 065013.

\bibitem[{Coll et~al.(2012)Coll, Ferrando, and
  Morales-Lladosa}]{coll2012positioning}
Coll, B., Ferrando, J.~J., Morales-Lladosa, J.~A., 2012. Positioning systems in
  minkowski space-time: bifurcation problem and observational data. Physical
  Review D 86~(8), 084036.

\bibitem[{Coll and Pozo(2006)}]{Coll:2006sg}
Coll, B., Pozo, J.~M., 2006. {Relativistic positioning systems: The Emission
  coordinates}. Class. Quant. Grav. 23, 7395--7416.

\bibitem[{Delva et~al.(2011)Delva, Kosti{\'c}, and
  {\v{C}}ade{\v{z}}}]{delva2011numerical}
Delva, P., Kosti{\'c}, U., {\v{C}}ade{\v{z}}, A., 2011. Numerical modeling of a
  global navigation satellite system in a general relativistic framework.
  Advances in Space Research 47~(2), 370--379.

\bibitem[{Gomboc et~al.(2013)Gomboc, Kosti{\'c}, Horvat, Carloni, and
  Delva}]{gomboc2013relativistic}
Gomboc, A., Kosti{\'c}, U., Horvat, M., Carloni, S., Delva, P., 2013.
  Relativistic positioning systems and gravitational perturbations. Acta Futura
  7, 79--85.

\bibitem[{Pascual-S{\'a}nchez et~al.(2004)Pascual-S{\'a}nchez, Miguel, and
  Vicente}]{pascual}
Pascual-S{\'a}nchez, J.-F., Miguel, A.~S., Vicente, F., 2004. Isotropy of the
  velocity of light and the sagnac effect. In: Rizzi, G., Ruggiero, M.~L.
  (Eds.), Relativity in Rotating Frames: Relativistic Physics in Rotating
  Reference Frames. Springer Netherlands, Dordrecht, pp. 167--178.

\bibitem[{Puchades et~al.(2021)Puchades, Arnau, and
  Fullana}]{colmenero2021relativistic}
Puchades, N., Arnau, J.~V., Fullana, M.~J., 2021. Relativistic positioning:
  including the influence of the gravitational action of the sun and the moon
  and the earth's oblateness on galileo satellites. Astrophysics and Space
  Science 366~(7), 1--19.

\bibitem[{Puchades and S\'aez(2014)}]{Puchades:2014oua}
Puchades, N., S\'aez, D., 2014. {Relativistic positioning: errors due to
  uncertainties in the satellite world lines}. Astrophys. Space Sci. 352,
  307--320.

\bibitem[{Rovelli(2002)}]{Rovelli:2001my}
Rovelli, C., 2002. {GPS observables in general relativity}. Phys. Rev. D 65,
  044017.

\bibitem[{Ruggiero et~al.(2011)Ruggiero, Capolongo, and
  Tartaglia}]{Ruggiero:2010nd}
Ruggiero, M.~L., Capolongo, E., Tartaglia, A., 2011. {Pulsars as celestial
  beacons to detect the motion of the Earth}. Int. J. Mod. Phys. D 20,
  1025--1038.

\bibitem[{Ruggiero and Tartaglia(2008)}]{Ruggiero:2007wh}
Ruggiero, M.~L., Tartaglia, A., 2008. {Mapping Cartesian coordinates into
  emission coordinates: Some toy models}. Int. J. Mod. Phys. D 17, 311--326.

\bibitem[{Synge(1960)}]{synge1960relativity}
Synge, J., 1960. Relativity: The General Theory. No. v. 1 in North-Holland
  series in physics. North-Holland Publishing Company.

\bibitem[{Tartaglia et~al.(2011{\natexlab{a}})Tartaglia, Ruggiero, and
  Capolongo}]{Tartaglia:2010sw}
Tartaglia, A., Ruggiero, M.~L., Capolongo, E., 2011{\natexlab{a}}. {A Null
  frame for spacetime positioning by means of pulsating sources}. Adv. Space
  Res. 47, 645--653.

\bibitem[{Tartaglia et~al.(2011{\natexlab{b}})Tartaglia, Ruggiero, and
  Capolongo}]{Tartaglia:2011ae}
Tartaglia, A., Ruggiero, M.~L., Capolongo, E., 2011{\natexlab{b}}. {A
  Relativistic navigation system for space}. Acta Futura 4, 33--40.

\end{thebibliography}

\end{document}